\documentclass[authoryear,12pt,3p,1in]{jowarticle}

\usepackage[english]{babel}
\usepackage[utf8]{inputenc}
\usepackage{amsmath,amsfonts,amssymb,amsthm}
\usepackage{graphicx}
\usepackage[colorinlistoftodos]{todonotes}
\usepackage{natbib}
\usepackage{blindtext}
\usepackage{subcaption}
\usepackage{caption}
\usepackage{setspace}
\usepackage{tikz-cd}
\usepackage{tensor}
\usepackage{accents}
\usepackage[shortlabels]{enumitem}
\newlength{\dhatheight}

\usepackage{newtxtext}
\usepackage{enumitem}
\setlist{nosep}

\newtheorem{thm}{Theorem}

\renewcommand{\footnotesize}{\normalsize}

\makeatletter
\renewcommand\section{\@startsection{section}{1}{\z@}{-3.25ex plus -1ex minus -.2ex}{1.5ex plus .2ex}{\normalsize\bf}}
\renewcommand\subsection{\@startsection{subsection}{2}{\z@}{-3.25ex plus -1ex minus -.2ex}{1.5ex plus .2ex}{\normalsize\bf}}
\renewcommand\subsubsection{\@startsection{subsubsection}{3}{\z@}{-3.25ex plus -1ex minus -.2ex}{1.5ex plus .2ex}{\normalsize\bf}}
\makeatother

\begin{document}
\begin{frontmatter}
\title{The Geometry of the ``Gauge Argument''}
\author{James Owen Weatherall}\ead{weatherj@uci.edu}
\address{Department of Logic and Philosophy of Science\\ University of California, Irvine}
\author{Benjamin H.~Feintzeig}\ead{bfeintze@uw.edu}
\address{Department of Philosophy\\ University of Washington, Seattle}
\begin{abstract}We show how the ``gauge argument'', widely criticized by philosophers as \emph{ad hoc} and unconvincing, may be reconstructed rigorously, and, we believe, convincingly, as an argument concerning the geometry of vector bundles.  In this form, the gauge argument has two parts, the first of which is strongly analogous to the Weyl-Stein argument for Galilean spacetime as the appropriate setting for Newtonian gravitation; and the second of which invokes a classic result in the Yang-Mills theory literature, due to Utiyama [1956, ``Invariant Theoretical Interpretation of Interaction'', \emph{Phys. Rev.} \textbf{101}, 1597].\end{abstract}
\end{frontmatter}

\pagenumbering{arabic}

\onehalfspacing

\section{Introduction}
The so-called ``gauge argument'', a mainstay of quantum field theory textbooks in which one ``derives'' the standard action for Yang-Mills fields interacting with charged matter from apparently very weak assumptions, has long been a source of controversy among philosophers of physics.  Physicists often claim that the gauge argument reveals the logic of nature, or that it explains the nature of observed interactions between matter and gauge fields \citep{Mandl+Shaw,Ryder,Gockeler+Schucker,Auyang,WeinbergQTF2}.  But on further scrutiny, philosophers of physics have concluded that the gauge argument is not convincing, at least as usually understood.  Some argue that it relies on steps that are ad hoc and unjustified. At best, they conclude, it may serve as a useful heuristic for constructing field theories \citep{TellerOld,Brown,Lyre,Martin,Martin2,Healey}.  Others claim that while the gauge argument does provide physical insight, its conclusion or significance is rather different from what is usually claimed \citep{Teller,Hetzroni}.\footnote{See \citet{Dougherty} for a recent review. \citet{Gomes+etal} offers yet another perspective on the gauge argument, though the argument they recover and discuss is quite different from the textbook argument criticized by other philosophers, and which we consider here.}

Here we will attempt to defend the gauge argument from these critics. We will show how the argument may be reconstructed, rigorously and convincingly, as an argument about the dynamics of a physical field represented by a section of a vector bundle, in the presence of plausible and clearly articulated assumptions about what features these dynamics should be expected to have.  The approach we take will have a distinctly geometrical character:  we will argue that the gauge argument is best construed as revealing the class of Lagrangians compatible with the geometry of vector bundles.\footnote{Though there are notable differences, the version of the argument we give is anticipated by \citet[\S 6.14]{Schutz}, though to our knowledge it has not previously been argued that the geometric perspective Schutz advocates provides the core of a response to the critiques from philosophers of physics.}

Perhaps the most striking---or at least, philosophically puzzling---feature of standard presentations of the gauge argument is that it seems to allow one to derive something from nothing.  In its usual guise, the starting point for the argument is ``gauge symmetry'', which is taken to be mere descriptive redundancy in the mathematical formalism of the theory.  The conclusion of the argument---the existence of interacting gauge fields and the fact that the interacting field is massless---is certainly physically significant.  One advantage of the presentation we give will be to thoroughly demystify this apparent magic.  The argument begins with substantive physical assumptions about the geometrical structure of our models, and it concludes that only some dynamics are compatible with that structure.  As we discuss below, in this form the argument has a familiar character: it is strongly analogous to the Weyl-Stein argument that the appropriate geometrical setting for classical Newtonian physics is Galilean spacetime, rather than Newtonian spacetime---and that the yet-weaker Leibnizian spacetime is too little structure to support the theory \citep{Weyl,Stein,EarmanWEST,WeatherallCST}.

From here, the argument will proceed as follows.  To fix ideas, in section \ref{GA} we will begin with a presentation of the gauge argument in its ``standard'' form---or at least, in the form that has been widely criticized by philosophers of physics.  In section \ref{critics}, we will present some of the problems that have been raised for the standard gauge argument, by the authors cited above.  Section \ref{GArevisited} will present the first part of our reconstruction of the gauge argument.  Then, in section \ref{Stein}, we will revisit the Weyl-Stein argument and note the strong structural similarity.  In section \ref{Utiyama} we will present the second part of our recovery, which runs via a classic result due to \citet{Utiyama} that establishes strong constraints on Lagrangians for principal connections compatible with a certain invariance principle.  We interpret this result as saying that the only (first-order, Lagrangian) dynamics for a derivative operator compatible with just the structure presupposed by matter dynamics must be a function of the curvature of the derivative operator.  We will conclude in section \ref{criticsRevisited} by revisiting the critics and evaluating what the argument, as we recover it, accomplishes.


\section{The Standard Presentation}\label{GA}

We will run the gauge argument for the example of a scalar field representing electrically charged matter interacting with electromagnetic fields.\footnote{This section follows closely the presentation given in \citet[pp. 223-4]{Martin}.  See also \citet[pp. 161-5]{Healey} and \citet[pp. 467-8]{Teller}.  We use a scalar field instead of a spinor field just to simplify the discussion, though nothing turns on this.  The same argument, and our recovery, works for other matter, including matter associated to non-Abelian frame bundles.}  Let $(M,\eta_{ab})$ be Minkowski spacetime.  (We work in signature $(+,-,-,-)$.)  Let $\varphi: M\rightarrow\mathbb{C}$ be a free field representing electrically charged matter.  The equation of motion for $\varphi$ is the (complex) Klein-Gordon equation, 
\[
\Box\varphi = -m^2\varphi\]
which is obtained by extremizing the following Lagrangian with respect to the field $\varphi$.
\begin{equation}\label{KGEq}
\mathcal{L}_{KG} (\varphi,\partial\varphi) = |\partial\varphi|^2 - m^2|\varphi|^2
\end{equation}

\noindent Now consider the transformation $\varphi(x)\mapsto\varphi'(x) = e^{i\Lambda}\varphi(x)$, for any constant $\Lambda\in\mathbb{R}$.  Notice that
\begin{equation*}
\mathcal{L}_{KG} (\varphi',\partial\varphi') = |\partial(e^{i\Lambda}\varphi)|^2 - m^2|e^{i\Lambda}\varphi|^2 = \mathcal{L}_{KG}(\varphi,\partial \varphi)
\end{equation*}

\noindent In other words, this transformation leaves the Lagrangian $\mathcal{L}_{KG}$ invariant, meaning that if the original field $\varphi$ satisfied the Euler-Lagrange equations, so will the new field $\varphi'$.  These transformations form a representation of the group $U(1)$ over the space of complex-valued functions $\varphi$.  They are called `global' gauge transformations because $\Lambda$ does not vary over spacetime.  The conserved Noether current associated with the global $U(1)$ symmetry is $J^a = i(\varphi^*\partial^a\varphi - \varphi\partial^a\varphi^*)$.  

These global gauge transformations are to be contrasted with \emph{local} gauge transformations, where we let $\Lambda$ vary over spacetime as a function $\Lambda: M\rightarrow\mathbb{R}$.  The Lagrangian $\mathcal{L}_{KG}$ is manifestly not invariant under local gauge transformations.
\begin{align*}
\mathcal{L}_{KG} (\varphi',\partial\varphi') &= |\partial(e^{i\Lambda}\varphi)|^2 - m^2|e^{i\Lambda}\varphi|^2
\\
&=|\partial \varphi|^2 + J^ad_a\Lambda + |\varphi|^2d_a\Lambda d^a\Lambda -m^2|\varphi|^2 \neq \mathcal{L}_{KG}(\varphi,
\partial\varphi')
\end{align*}

\noindent But a small change to the Lagrangian restores invariance.  Let
\begin{equation*}
\mathcal{L}_{KG+}(\varphi,\partial\varphi,A_a) = \left[(\partial_a + iA_a)\varphi\right]\left[(\partial^a - iA^a)\varphi^*\right] -m^2|\varphi|^2
\end{equation*}

\noindent where we have introduced a new covector field $A_a$ on $M$, and written out the kinetic term in greater detail.  Consider local gauge transformations for any $\Lambda(x)$ of the form $\varphi\mapsto \varphi'(x) = e^{i\Lambda(x)}\varphi(x)$ and $A_a(x) \mapsto A_a'(x) = A_a(x) - \partial_a\Lambda(x)$.  We find the Lagrangian is invariant under these new local gauge transformations since,
\begin{equation*}
\begin{split}
\mathcal{L}_{KG+}(\varphi',\partial\varphi',A_a')&= \mathcal{L}_{KG}(\varphi',\partial\varphi') - \tilde{J}{}^a(A_a-\partial_a\Lambda(x))\\
&= \mathcal{L}_{KG}(\varphi,\partial \varphi) - \tilde{J}{}^a(\partial_a\Lambda(x))  - \tilde{J}{}^a A_a + \tilde{J}{}^a(\partial_a \Lambda(x))\\
&= \mathcal{L}_{KG}(\varphi,\partial \varphi) - \tilde{J}{}^a A_a = \mathcal{L}_{KG+}(\varphi,\partial\varphi,A_a),
\end{split}
\end{equation*}
\noindent where we have defined $\tilde{J}{}^a=i(\varphi^*(\partial^a+iA_a)\varphi - \varphi(\partial^a-iA_a)\varphi^*)$.

Writing the Lagrangian in the way we just have is highly suggestive.  It appears we have a new field $A_a$, and a new interaction -- that is, a new coupling in the Lagrangian -- between this field and (a modified form of) the Noether current associated with our initial global $U(1)$ symmetry.  But one thing is not quite in order yet.  This new field has a potential term, given by its interaction with $\varphi$.  But it does not have a kinetic term.  This means that when we vary the Lagrangian with respect to this new field, we do not get standard equations of motion. To fix this, we need to add one more term to the Lagrangian, related to the derivative of this new field.  To be consistent with the foregoing, we should choose this term to be invariant under local gauge transformations.  Define $F_{ab} = \partial_a A_b - \partial_b A_a$.  Now we have the final Lagrangian:
\begin{equation*}
\mathcal{L}_{final}(\varphi,A_a) = \mathcal{L}_{KG+}(\varphi,A_a) - \frac{1}{4} F_{ab}F^{ab}
\end{equation*}

\noindent One can confirm that this Lagrangian is still invariant under local gauge transformations, as required.

But now it seems a miracle has occured.  Varying $\mathcal{L}_{final}$ with respect to $A_a$ yields not only sensible equations of motion, but very familiar ones: they are the equations of motion for the electromagnetic field, namely Maxwell's equations, sourced by $\tilde{J}{}^a$.  And if we vary with respect to $\varphi$, we find the equations of motion for a (charged) Klein-Gordon field in a background electromagnetic field.  In other words, we have derived electromagnetism by beginning with a certain kind of matter and imposing a new symmetry principle.  At this point, the standard account goes, we declare victory. 

Actually, there is a bonus worth mentioning. This process was constrained by the requirement that local gauge invariance be preserved.  Maintaining that requirement means that other possible terms combining $A_a$ and $\varphi$ are excluded from consideration.  For instance, the usual way of adding a mass term for the field $A_a$ would be to add a term of the form $\frac{1}{2} m_{photon}^2A^a A_a$.  But such a term is not invariant under local gauge transformation, so the particle corresponding to the field $A_a$, which is the photon, must have zero mass.

\section{Peanuts from the Gallery}\label{critics}

\citet{Martin} notices three problems with the gauge argument as just presented. First, he argues, when scrutinized, the usual ways of justifying the move from global to local gauge invariance fail. (We did not attempt to justify this step.) Second, imposing the requirement of local gauge invariance does not uniquely dictate the form of the Lagrangian---there are many other possible gauge invariant terms we might add, and we only get a unique Lagrangian when we take into account other physical requirements such as Lorentz invariance and renormalizability. And finally, no step of the gauge argument necessitates that we allow $F_{ab}$ to be nonzero.  After all, we started with the Lagrangian for a free field and so, we might expect $A_a$ to always be gauge transformable to the zero field.  It seems we have not derived the existence of a new field $F_{ab}$ at all.  Instead, we have put it in by hand.

\subsection{Why go local?}

Martin considers two ways to justify the move from global to local gauge invariance, both based on arguments he detects in the physics literature.  First, one might try to argue that local gauge invariance comes along with being a local field theory.  But if by local field theory one means, as we usually do, that the Lagrangian at a point depends only on the field and a finite number of derivatives at that point, then this has nothing to do with gauge invariance because the initial Lagrangian $\mathcal{L}_{KG}$ already satisfies this requirement \citep[p. 225]{Martin}.

Second, one might argue that local gauge invariance comes from constraints placed on locality by the theory of relativity. For instance, one might say that global gauge transformations are a way of enacting the same change on our fields everywhere at the same time, which is ruled out by relativity (the argument goes) because one could not communicate how to perform the transformation to all locations simultaneously.  But notice that neither local nor global gauge transformations require any instantaneous communication at a distance; gauge transformations are passive, like changing coordinates, which means that we do not think that performing a gauge transformation makes any physical difference to the fields (Martin 2002, p. 226; Healey 2007, p. 162).  So there is nothing to communicate. In any case, the global gauge transformations form a subgroup of the local gauge transformations so by requiring local gauge invariance one also requires global gauge invariance. If global gauge transformations are a problem, so too are local ones \citep[p. 226]{Martin}.

\subsection{Nonuniqueness}

Although it may appear that imposing local gauge invariance allowed us to derive the Lagrangian, Martin argues that it does not uniquely lead to the final Lagrangian $\mathcal{L}_{final}$.  One might add other gauge invariant terms such as a Pauli term, which is proportional to $(\partial_a+iqA_a)\varphi^*(\partial_b+iqA_b)\varphi F^{ab}$.  This would maintain both the gauge invariance and Lorentz invariance of the Lagrangian.  One can, however, uniquely specify the Lagrangian by taking into account further considerations besides gauge invariance.  The minimally-coupled Lagrangian we arrived at is ``the simplest, renormalizable, Lorentz and gauge invariant Lagrangian yielding second order equations of motion for the coupled system" \citep[p. 228]{Martin}.  So if one requires a certain kind of formal simplicity, renormalizability, and Lorentz invariance as well, then the gauge argument gives us exactly what we wanted.  But then it is no longer clear that local gauge invariance is doing any of the work---it might be the renormalizability requirement that really constrains the form of the Lagrangian.  This undermines the idea that all we started with was the requirement of gauge invariance and that that dictated the form of our theory.

\subsection{Reality of the fields}

Some say that the gauge argument leads us to the very existence of interacting gauge fields, but the treatment of $A_a$ as a physical field ``is ultimately put in by hand" \citep[p. 229]{Martin}.  We started with the Lagrangian $\mathcal{L}_{KG}$ for the \emph{free} field $\varphi$.  Imposing the requirement of local gauge invariance led us to the Lagrangian $\mathcal{L}_{KG+}$.  But if we expect this to be the Lagrangian for the same free field, then we would expect $A_a$ to be gauge-equivalent to the zero field.  Consequently, we would expect $F_{ab}$ to be the zero field.  And if we expect the gauge field $F_{ab}$ to always be zero, then we seem to lose our motivation for adding the kinetic term to the Lagrangian in the first place.  It will also always be zero.

On the other hand, adding the kinetic term and allowing $F_{ab}$ to be nonzero amounts to treating the new fields as physical rather than merely formal artifacts.  But nothing necessitates this treatment of the fields as physical.  The motivation comes from some other source, such as a formal analogy with electromagnetism, since the field $A_a$ transforms under gauge transformations in the same way as the vector potential in electromagnetism \citep[p. 229]{Martin}; \citep[cf. ][p. 165]{Healey}.  But local gauge invariance does not force this identification.  More, it seems all the work is done by requiring the field $A_a$ to be like a Maxwell vector potential, and so it is hardly a surprise that in the end it satisfies Maxwell's equations.  Treating the new fields as physical is another, and seemingly independent, input to the gauge argument.

\section{The Gauge Argument Revisited}\label{GArevisited}

The concerns presented in the previous section have merit---at least in the context of the argument as presented in section \ref{GA}.  But if the gauge argument is bankrupt, why does it play such a powerful heuristic role in constructing physical theories?  We claim that although the argument as presented has a number of shortcomings, at its core there are principled considerations with real physical---and mathematical---significance.  In this section, we will attempt to draw them out.  The result will be a new presentation of what we take to be substantially the same argument, but with clearer assumptions and a more precisely articulated conclusion.

\subsection{Preliminary Remarks}

To begin, let us return to the starting point of the gauge argument, as we presented it in section \ref{GA}.  The very first step, recall, was to write down a Lagrangian from which one could obtain the (charged) Klein-Gordon equation:
\begin{equation}\tag{\ref{KGEq}}
\mathcal{L}_{KG} (\varphi,\partial\varphi) = \frac{1}{2}|\partial\varphi|^2 - \frac{1}{2}m^2|\varphi|^2
\end{equation}
Following Martin and others, we quickly moved on to manipulating this Lagrangian, without further comment.\footnote{As another matter of notation, which should have been clarified sooner except that it would have ruined the narrative tension of the paper: throughout, we consider first-order Lagrangian densities, which we understand as functions on the first jet bundle over a bundle of fields.  The arguments of the Lagrangian density will be written uniformly as $\varphi,\partial\varphi$ to reflect that the Lagrangian depends on field values and field value jets.  We should not interpret that $\partial$ symbol her as referring to a particular derivative operator; rather, it is a 1-jet, and the expression on the right-hand-side, which does depend on a derivative operator, depends on the 1-jet and derivative operator together in the standard way.}  But in fact, quite a bit has gone into writing this Lagrangian down.  For one, the Lagrangian makes reference to a ``free Klein-Gordon field'' $\varphi$ on a spacetime manifold $M$.  This field is a complex scalar field, which means that it is not a ``tangent field'' in the ordinary sense of the (real) manifolds we use to represent spacetime \citep{March+Weatherall}.  Instead, we defined it in section \ref{GA} as a smooth map from $M$ to $\mathbb{C}$.  

Defining a complex scalar field in this way may seem natural, given the apparent analogy to real scalar fields.  But on reflection, this analogy is weaker than it initially seems.  The reason is that the real numbers have a rigidity property that the complex numbers (and other, higher dimensional real and complex vector spaces) lack. Any two complete ordered fields (i.e., any two copies of the real numbers) are \emph{uniquely} isomorphic.  This, in turn, has several consequences, including that there is a privileged standard of differentiation for real-valued fields on any real manifold (namely, the differential operator $d$) and that every diffeomorphism lifts uniquely to act on real scalar fields without ambiguity \citep[c.f.][\S 3]{March+Weatherall}.  Neither of these is true for the complex numbers.  

Reflecting on these facts suggests a different perspective on the field $\varphi$.  Rather than think of it as a map from $M$ to $\mathbb{C}$, one might instead understand it as a section of some vector bundle $\pi:E\rightarrow M$, whose typical fiber is the one dimensional complex vector space $\mathbb{C}$.\footnote{For further details on fiber bundles, including vector bundles, see \citet{Kobayashi+Nomizu}; for applications to physics see especially \citet{Bleecker} and \citet{Palais}.  Here we draw heavily on the treatment given by \citet{WeatherallFBYMGR}.}  In that language, our initial assumption that $\varphi$ can be seen as a map to $\mathbb{C}$ amounts to the requirements that (a) $E$ be a \emph{trivial bundle}, i.e., one isomorphic, as bundles, to the product manifold $M\times \mathbb{C}$; and (b) that a global trivialization $\zeta:E\rightarrow M\times\mathbb{C}$ is fixed.  With that structure, every section of $E$ becomes a map to $\mathbb{C}$, and vice versa.  All we have done here is adopt a somewhat different vernacular, without changing any substance.  We have simply highlighted that if we think in terms of vector bundles, our definition of a complex scalar field fixes a global trivialization of a trivial bundle; and that, in disanalogy with the real scalar field case, we could have chosen a different global trivialization.\footnote{\label{fn:real}Note that the field structure and ordering on the reals are essential to this claim.  See the paragraph to come and fn. \ref{fn:field}.}

We have also implicitly assumed that $\mathbb{C}$, or rather, the fibers of $E$, has some additional structure: when we write down expressions such as $|\varphi|^2$, we are assuming that there is an inner product available, relative to which we can take scalar products of $\varphi$ and its complex conjugate.  Of course, there is a canonical Hermitian inner product on $\mathbb{C}$, unique up to isomorphism, and so there is no real ambiguity in the expression as written.\footnote{\label{fn:field}This is the core of the disanalogy with real scalar fields alluded to above and in fn. \ref{fn:real}.  Even with the inner product, there is not a unique global trivialization for a complex vector bundle, because the field lacks an order.} But the compact form we have chosen suppresses the fact that we are making use of that structure. It also suppresses the fact that we are actually taking two \emph{different} inner products in the first term of the Lagrangian, since the derivatives $\partial$ define $\mathbb{C}$-valued one forms on $M$, and we are taking their product using the space-time metric.

Thus, a more explicit---albeit pedantic---way of writing the same Lagrangian, emphasizing the various products, would be as follows.
\begin{equation}\label{eq:KGLfinal}
\mathcal{L}_{KG} (\varphi,\partial\varphi) = \frac{1}{2}\eta^{ab}k_{AB}\partial_a\varphi^A\partial_b\varphi{}^B - \frac{1}{2}m^2k_{AB}\varphi^A\varphi{}^B
\end{equation}
Here we have used lower-case Latin characters to represent tangent space indices, and upper-case Latin characters to represent $E$ indices.\footnote{For more on this notation, see \citet{WeatherallFBYMGR}.}  The field $\eta^{ab}$ is the (Minkowski) spacetime metric, whereas $k_{AB}$ is the Hermitian inner product on each fiber $\mathbb{C}$.  (Here we have dropped any explicit notation to mark complex conjugation, because that operation is built into the Hermitian inner product $k_{AB}$.)

There is one more feature of this Lagrangian worth emphasizing in light of the foregoing.  Note that it refers not just to the field $\varphi$, but also to a \emph{derivative} of this field, $\partial_a\varphi$.  But where does this derivative operator $\partial$ come from?  It cannot be the spacetime derivative operator determined by $g_{ab}$: that derivative operator acts on sections of tangent bundle; it bears no relation at all to the vector bundle $E$ \citep{March+Weatherall}.  Indeed, without further structure, there is no privileged notion of differentiation on a general vector bundle. But in fact, there is no ambiguity, or rather, we have already implicitly addressed this issue when we stipulated that our bundle $E$ comes equipped with a fixed global trivialization $\zeta$, because any global trivialization determines a derivative operator, analogously to how a coordinate chart determines a derivative operator on a manifold.  In general, the derivative operator depends on that choice, in the sense that a change of trivialization, implemented, for instance, by some vertical vector bundle automorphism $\Psi:E\rightarrow E$, determines a \emph{different} derivative operator.  This is true even if we limit attention to derivative operators relative to which the inner product $k_{AB}$ is constant, and vertical vector bundle automorphisms that preserve the inner product.\footnote{Note that in some cases where we suppose there is additional structure on the fibers, there \emph{is} a unique derivative operator compatible with that structure. (Recall the fundamental theorem of Riemannian geometry.) But this is not true in every case, including the present one.}

\subsection{The Geometric Gauge Argument}

With those remarks in mind, we now return to the gauge argument proper---or rather, our geometrized version of it.  The starting point for the argument is a question implicit in our discussion of fiber bundles and global trivializations. As we have set things up, complex scalar fields are sections of a complex vector bundle with a fixed global trivialization.  This is unusual from a bundle perspective---it is analogous to working with a fixed global coordinate system on a smooth manifold. Much of the power of the fiber bundle formalism comes from dropping any preferred trivialization.  Do we \emph{need} to fix a global trivialization to write down the Lagrangian in Eq. \eqref{eq:KGLfinal}?  

In the first instance, the answer to this question is ``yes'', for reasons related to our discussions of derivative operators above.  Suppose we wrote $\mathcal{L}_{KG}$ relative to a different global trivialization, $\tilde{\zeta}:E\rightarrow M\times\mathbb{C}$.  Recall that the role that $\zeta$ played was to fix the derivative operator $\partial$. In general, a different trivialization would determine a different derivative operator $\tilde{\partial}$.  We find that $\partial=\tilde{\partial}$ (if and) only if the two trivializations are relatively constant, i.e., they differ by a global constant phase (see below).  So, for a general transformation, we find:
\[
\tilde{\mathcal{L}}_{KG}(\varphi,\partial\varphi) = \frac{1}{2}\eta^{ab}k_{AB}\tilde{\partial}_a\varphi^A\tilde{\partial}_b\varphi^B - \frac{1}{2}m^2k_{AB}\varphi^A\varphi^B\neq \mathcal{L}_{KG}(\varphi,\partial\varphi)
\]
How do we know $\tilde{\mathcal{L}}_{KG}\neq \mathcal{L}_{KG}$?  We can verify it explicitly, using the fact that the different derivative operators one may define on $E$ are systemically related to one another.  In particular, given any pair of derivative operators---say, $\partial$ and $\tilde{\partial}$---there exists a unique field $iA_a$ that allows one to express the action of the one operator in terms of the other, as follows: given any section $\varphi^A:M\rightarrow E$, $\partial_a\varphi^A = (\tilde{\partial}_a + iA_a)\varphi^A$; likewise $\partial_a\varphi_A=(\tilde{\partial}_a - iA_a)\varphi_A$, where here we write $\varphi_A=k_{AB}\varphi^B$.\footnote{We use $k_{AB}$ to lower the index here to make the notation for the partial derivative, and particularly the sign of $iA_a$, look more familiar.}  (More generally, if we were working with higher-dimensional vector spaces, the same would be true, except that $A_a$ would be replaced by an endomorphism valued form, $A^A{}_{Ba}$.\footnote{In fact, $A_a$ is also an endomorphism-valued form.  But since endomorphisms on $\mathbb{C}$ are complex numbers, we suppress the indices and just allow the endomorphism to act as scalar multiplication.})  

So, rewriting our original Lagrangian in terms of $\tilde{\partial}$, we find:
\begin{equation}\label{eq:transformed}
\mathcal{L}_{KG} (\varphi,\partial\varphi) = \frac{1}{2}\eta^{ab}(\tilde{\partial}_a + iA_a)\varphi^A(\tilde{\partial}_b - iA_b)\varphi_A - \frac{1}{2}m^2\varphi^A\varphi{}_A \neq \tilde{\mathcal{L}}_{KG}(\varphi,\partial\varphi)
\end{equation}
This verifies that the transformed Lagrangian does not equal the original one, unless $A_a$ vanishes, as in the relatively constant case mentioned above.  But it also highlights why not: what we see here is that the Lagrangian depends on a choice of derivative operator.  Transformations that preserve the derivative operator leave the Lagrangian invariant, whereas those that vary the derivative operator yield different Lagrangians.  On the other hand, we also see an ambiguity in what it means to write the Lagrangian relative to a choice of global trivialization.  If we understand the derivative operator to be fixed, we can always express that fixed derivative operator relative to different choices of global trivialization.  That is another sense of ``changing global trivialization''.  It shows a sense in which the Lagrangian does \emph{not} depend on a global trivialization after all---instead, it depends on a choice of derivative operator.  Indeed, if we simply stipulate a derivative operator from the start, and understand $\varphi$ as a section of a vector bundle, we need not introduce a global trivialization at all.

Now observe that the Lagrangian in Eq. \eqref{eq:transformed} is identical to the ``transformed'' Lagrangian above that was interpreted as including an ``interacting'' term.  But this is not a ``transformed'' Lagrangian at all!  Instead, it is exactly the \emph{same} Lagrangian, expressed relative to a different choice of derivative operator---or equivalently, relative to a different choice of global trivialization.  Thus, the ``interacting'' term does not represent an interaction at all; it represents the freedom to express any derivative one likes in terms of any other.

Here is another way of putting the point that makes the relationship with the gauge argument as formulated above more immediate.  (Here we suppress the $A$ indices for convenience.)  Suppose that we have chosen some Klein-Gordon field $\varphi$, which we now recognize as a (global) section, $\varphi:M\rightarrow E$, of a vector bundle $\pi:E\rightarrow M$.  Suppose, too, that we have fixed some (global) trivialization of $E$, and that $\partial$ is the derivative operator determined by this trivialization.  The particular field configuration $\varphi$ may be associated with an action on any region $O\subseteq M$, where this action is determined by integrating the scalar density $\mathcal{L}$ over $O$ (relative to a volume form determined by $\eta_{ab}$), where $\mathcal{L}$ is written using $\partial$.

Now suppose we consider some vertical vector bundle automorphism, $\Phi:E\rightarrow E$, that preserves the inner product $k_{AB}$.  Composing this map with $\varphi$ will yield a new section of $E$, $\Phi\circ \varphi:M\rightarrow E$---which, in turn, may be thought of as a new field configuration.  But the map $\Phi$ is naturally construed as an isomorphism of the inner product bundle we are considering: it preserves the vector space structure, the inner product, and even the points over which the fibers are defined. It should follow that any physical situation that could be represented by $\varphi$ could be equally-well represented by $\Phi\circ\varphi$, since after all, $\pi:E\rightarrow M$ is equivalent to $\pi:\Phi[E]\rightarrow M$ as vector bundles, and $\Phi\circ\varphi$ is the section of the latter corresponding to $\varphi$.

This last point was made in a strange way, however.  After all, we specified initially that $\Phi$ is an \emph{automorphism} of $E$.  Thus, one can think of $\varphi$ and $\Phi\circ\varphi$ as two sections of the same bundle---that is, as two distinct configurations of the same field that happen to be related in a certain way.  This means, in particular, that the Lagrangian density $\mathcal{L}_{KG}$, and the associated action over any region, may be computed for the field $\Phi\circ\varphi$. What is the relationship between the Lagrangian densities associated with these two fields?  As it turns out, they are the same---though to see this requires some careful attention to the action of $\Psi$ on the various objects appearing in $\mathcal{L}_{KG}$.  First, observe that there is a more transparent way of expressing $\Phi\circ\varphi$: namely, because $\Phi$ preserves the inner product on each fiber of $E$, it must take $\varphi$ at each point to another vector at that point of the same length.  This means that there must be a (real) scalar field $\Lambda$ on $M$ such that $\Phi\circ\varphi = e^{i\Lambda}\varphi$.  This leads us to the following equation, where we now reintroduce indices for precision:
\[
\mathcal{L}_{KG} (\Phi\circ\varphi,\partial(\Phi\circ\varphi)) = \frac{1}{2}\eta^{ab}\tilde{\partial}_a\left(e^{i\Lambda}\varphi^A \right)\tilde{\partial}_b\left(e^{-i\Lambda}\varphi_A \right) - \frac{1}{2}m^2\varphi^A\varphi_A
\]

Note that in this equation, we have written $\tilde{\partial}$ instead of $\partial$.  The reason is that the sense in which $\varphi$ and $\varphi e^{i\Lambda}$ should be taken as ``equivalent'' is given by the transformation $\Phi$.  But $\partial$, as we noted above, is a derivative operator determined relative to a particular choice of global trivialization, $\zeta:E\rightarrow M\times\mathbb{C}$.  The transformation $\Phi$ generates a new global trivialization: $\zeta\circ \Phi:E\rightarrow M\times \mathbb{C}$.  Like any trivialization, this new one gives rise to a standard of differentiation, coinciding with $\partial$ only if $\Lambda$ is a global constant.  It is this derivative operator that we have written as $\tilde{\partial}$.  And as we have now emphasized several times, these are systemically related: in fact, given any section $\varphi:M\rightarrow E$, we have $\tilde{\partial}_a\varphi^A = (\partial_a - i \partial_a\Lambda) \varphi^A$ and $\tilde{\partial}_a\varphi_A = (\partial_a + i\partial_a\Lambda)\varphi_A$, where $\Lambda$ is precisely the (real) scalar field implementing the transformation $\Phi$ on sections above.

The discussion so far trades on a basic fact about derivative operators on a vector bundle: one can freely express any one in terms of any other.  It is the field $A_a=\partial_a\Lambda$ that arises when we exercise this freedom that, in the standard gauge argument, is interpreted as dictating the ``form'' of a new sort of interaction.  In a sense, though, these remarks only put further pressure on the question we asked above.  The $A_a$ fields, we now see, arise from expressing one derivative operator in terms of another.  But this tells us nothing about which derivative operator we should have used in the first place.  It is at this point that substantive new physics enters.  We make the following Ansatz: suppose that the derivative operator appearing in the Lagrangian---which we will now write as $\nabla$, since we no longer suppose it is determined by any trivialization---is \emph{itself} a dynamical quantity, dependent on the configuration $\varphi^A$.  One way to motivate this choice would be by appealing to an analogy with general relativity: in that case, too, one needs to choose which derivative operator will appear in the dynamics of fields on spacetime, and it turns out that the choice is determined by Einstein's equation.  Another way to motivate the choice would be to point out that if the derivative operator is not determined by some physical principle, then it seems like an \emph{ad hoc} choice.  But we emphasize that these are merely heuristic motivations.  In the end, the decision to treat the derivative operator as a dynamical quantity is itself a new hypothesis, in no way fixed by what has come before.\footnote{In recent work, \citet{Bhalla-Ladd} has offered an argument, highly sympathetic with our approach here, that in fact a dynamical connection is required by gauge naturality.}   (This step plays the role of the decision above to make $A_a$ ``real'' by introducing a ``kinetic term''---though now it is motivated by quite different considerations.)

What \emph{is} fixed however, or at least constrained, is the character of the dynamical principle by which the derivative operator will be determined.  In particular, we suppose that the hypothesis above is implemented by introducing into the Lagrangian a new term that depends only on the (covariant) derivative operator and other structure already introduced, such as the spacetime metric or Killing metric, and not on any particular representation of the derivative operator or other new structure.  This condition may be understood as the requirement that the additional term in the Lagrangian depends only on the vector bundle structure already defined---that is, the vector bundle, the inner product on each fiber, and now a choice of derivative operator---and not on any further structure, such as a global trivialization; more formally, we may understand it as the requirement that the additional term be invariant under the same vertical vector bundle isomorphisms that already leave the Lagrangian density $\mathcal{L}_{KG}$ invariant---again, once sufficient care is paid to the action of such transformations on derivative operators.

It turns out that there are \emph{many} such terms that could be added.  But as we will make clear below, there is a precise sense in which they all have a shared form: they are all scalar functions of the \emph{curvature} of the derivative operator, which is a two form $F_{ab}$.\footnote{Technically, this is an endomorphism-valued two-form, defined so that for any section $\varphi^A$, $\nabla_{[a}\nabla_{b]}\varphi^A = F^A{}_{Bab}\varphi^B$, but again the endomorphisms are just complex numbers.}  For instance, one particularly simple (non-vanishing) such function is of the form $k F^{ab}F_{ab}$, where spacetime indices are raised with the inverse spacetime metric and $k$ is some constant.  And indeed, we recognize this term as precisely the ``kinetic'' term added to the Klein-Gordon Lagrangian above, in the standard version of the gauge argument.  This term is not the unique Lagrangian that could be associated with a derivative operator and invariant under the transformations discussed; it is, however, the unique such term that is \emph{quadratic} in the curvature and depends only on the structure already introduced (namely, the derivative operator and the spacetime metric).  In this sense, it is the unique term that is both invariant under those transformations and having the form of a kinetic term in a Lagrangian.

\section{The Weyl-Stein Argument}\label{Stein}

Before proceeding to discuss derivative operator dynamics in more detail, we will pause to reflect on the structure of the argument we have just given.  As we have recovered it, the core of the gauge argument concerns the structure needed to express the dynamics of matter.  In this form, the argument looks strikingly similar to another one, beloved of philosophers of physics: the Weyl-Stein argument for Galilean spacetime over Newtonian spacetime, on the grounds that Newtonian spacetime posits more structure than is needed to support Newtonian particle dynamics \citep{Weyl,Stein,EarmanWEST,WeatherallCST}.\footnote{Much of what we say here could be recast in a slightly different way, emphasizing the relationship to Earman's famous symmetry matching conditions (SP1) and (SP2) \citep{EarmanWEST,Weatherall+MarchNT}.  But we leave the details to the reader.}

Recall how this argument goes.  Anachronistically: Newton begins by supposing he needs a structure $(M, t_a, h^{ab}, \nabla, \xi^a)$ (``Newtonian spacetime'') to express his laws.  (Here $M$ is a smooth manifold, $t_a$ is a ``temporal metric'' on $M$, $h^{ab}$ is a ``spatial metric'' on $M$ satisfying $h^{ab}t_b = \mathbf{0}$, $\nabla$ is a covariant derivative operator satisfying $\nabla_a t_b = \mathbf{0}$ and $\nabla_a h^{bc} = \mathbf{0}$, and $\xi^a$ is a vector field satisfying $\xi^a t_a = 1$ and $\nabla_a \xi^b = \mathbf{0}$.)  Newtonian spacetime posits an affine structure, a space/time split, and, crucially, a preferred standard of rest, given by $\xi^a$, that provides the resources to describe motion relative to ``absolute space''.  But as philosophers from Leibniz on have noted, often on the basis of symmetry arguments, in fact Newon's laws of motion require less than this.  That is, as Newton himself noted in Corollary V to the laws of motion, Newtonian dynamics do not depend on the notion of absolute rest or absolute velocity.  We now know that it suffices to posit absolute \emph{acceleration}, while only relative velocity is needed.  (It seems neither Newton nor Leibniz understood that absolute velocity and absolute acceleration could come apart \citep{Stein}.)  We can see this by observing that the dynamics are invariant under \emph{Galilean transformations}, which are diffeomorphisms $\psi:M\rightarrow M$ that preserve $t_a$, $h^{ab}$, and $\nabla$, but may not preserve $\xi^a$.  It follows that Newtonian dynamics requires, at most, the structure preserved by the Galilean transformations, $(M, t_a, h^{ab}, \nabla)$ (``Galilean spacetime''). 

Galilean spacetime has ``less'' structure than Newtonian spacetime \citep{EarmanWEST,barrett2015spacetime,WeatherallCategories,WeatherallCST}. This can be seen in multiple ways.  Intuitively, a structure $(M, t_a, h^{ab}, \nabla, \xi^a)$ includes more things than $(M, t_a, h^{ab}, \nabla)$ (viz., an extra field that is not implied by the other structures). Alternatively, and more precisely, the symmetry group of Newtonian spacetime is a proper subgroup of that of Galilean spacetime. Thus, Newtonian physics presupposes less structure than Newton assumed. Several authors have also considered whether one could get away with even less structure, especially in light of Corollary VI to the Laws of Motion in Newton's Principia, which shows that Newtonian dynamics is invariant under additional transformations that map between unaccelerated and uniformly accelerated trajectories \citep{DiSalle,SaundersNewton,Knox,WeatherallNCKS,Dewar,Chen,March}.  Whether this is possible, in a fully satisfactory way, is controversial. But it is clear that some structures have too \emph{little} structure.  For instance, suppose we considered a diffeomorphism $\psi:M\rightarrow M$ that preserved only the still-weaker structure $(M, t_a, h^{ab})$ (``Leibnizian spacetime''). This map will not generally take solutions to solutions: e.g., $F^a = m\gamma^n \nabla_n \gamma^a$ does not imply that $\psi_*(F^a) = m(\psi\circ\gamma)^n \nabla_n (\psi\circ\gamma)^n$.  Nonetheless, the following always holds: given any $\psi:M\rightarrow M$, there exists a smooth field $C^a{}_{bc}$ on $M$ such that if $F^a = m\gamma^n \nabla_n \gamma^a$, then $\psi_*(F^a) = m(\psi\circ\gamma)^n \nabla_n (\psi\circ\gamma)^n - m C^a{}_{mn}(\psi\circ\gamma)^m(\psi\circ\gamma)^n$. This is because there exists some $C^a{}_{bc}$ such that $\psi_*(\nabla) = (\nabla, C^a{}_{bc})$. That $C^a{}_{bc}$ may be non-zero for this class of transformations is a way of expressing the idea that Newton's laws require a derivative operator $\nabla$.  

Now consider again the argument in the previous section.  Structurally, it is identical to the one just rehearsed.   We begin by positing one kind of structure, which we use to write down dynamical laws.  That structure is something we might write as $(M,g_{ab},E,k_{AB},\zeta,\partial)$, where $(M,g_{ab})$ is our spacetime, $E$ is a vector bundle over $M$, $k_{AB}$ is a metric on the fibers of $E$, and $\zeta$ is a global trivialization with associated derivative operator $\partial$.  We then find that not all of this structure is needed to express the laws.  In fact, the only role $\zeta$ plays is to determine a derivative operator; any transformation that preserves that derivative will be a symmetry of the laws.  These are the ``global'' gauge transformation, i.e., the ones that are constant relative to $\zeta$'s derivative operator.  This indicates that the laws require only a weaker structure---namely, one that does not impose a fixed global trivialization.  That is, we only need $(M,g_{ab},E,k_{AB},\partial)$, for some choice of $\partial$.  This is less structure than we began with, but it is sufficient to express the Lagrangian.

Can we get away with even less?  To answer that, we consider yet a broader class of transformations, namely local gauge transformations, which are analogous, now, to Leibnizian transformations.  These are transformations that preserve only $(M,g_{ab},E,k_{AB})$.  When we do this, we find that the laws are no longer invariant.   To ``restore'' invariance, we must reintroduce the derivative operator as part of the structure, and then push it forward along the transformation.  This shows that, precisely as with Galilean spacetime, the laws depend on a derivative operator, but no more. That is where the argument of the previous section concluded: Klein-Gordon theory requires a structure analogous to Galilean spacetime, intermediate between one with a fixed global trivialization and one that posits a bare vector bundle.\footnote{In fact, this analogy is even stronger than it seems.  A classical spacetime can be thought of an affine bundle over the manifold $\mathbb{R}$, with fibers consisting of three dimensional spacelike surfaces.  This structure is endowed with a metric $t_a$ on the base space and a positive definite metric $h_{AB}$ on each fiber.  Newtonian spacetime adds to this structure a global trivialization, identifying points in different fibers with a fixed copy of the manifold $\mathbb{R}^3$, whereas Galilean spacetime adds only a connection that is compatible with the metrics.  So in both cases, we start by assuming a global trivialization, and conclude that a derivative operator suffices.} And at that stage, the question arises what status that derivative operator has.  In the classical Weyl-Stein argument, this was the stopping point: a fixed derivative operator is simply posited.  That is an option in the present case as well.  Another option, following Newton-Cartan theory or general relativity, would be to suppose that that derivative operator itself be determined dynamically.  That is the option standardly pursued in the final step of the gauge argument---and the one to which we now return.

\section{Connnection Dynamics}\label{Utiyama}

In section \ref{GArevisited}, we claimed that the Lagrangians suitable for describing the dynamics of a derivative operator on the Klein-Gordon bundle are strongly constrained by the geometry of the vector bundle they act on, and that, with some mild compatibility requirements, the only possible Lagrangians are scalar functions of the curvature of the derivative operator.  This result is originally due to \citet{Utiyama}, though the way the theorem was first stated obscures its geometrical interpretation; the presentation we give here most closely follows that of \citet{Bleecker}.  Of course, the result itself is well-known, and in a sense underlies the standard presentation of the gauge argument as well.  Our goal here is to show why the assumptions underlying the theorem are naturally interpreted in the ways described above; and to precisely characterize the class of Lagrangians that are compatible with these constraints.  We begin with some preliminaries, and then state the key result and explain how it bears on our argument.

\subsection{Preliminaries: First-Order Lagrangians for Connections}

In order to characterize the Lagrangian densities for derivative operators compatible with certain constraints, we first need to say in more detail what we mean by a Lagrangian density in the first place, and also characterize the space of possible derivative operators on a vector bundle.  To describe Lagrangians, it will be convenient to work with the theory of jet bundles, which permits a general characterization of a space of fields and their derivatives.\footnote{For background on the theory of jet bundles, see \citet{Saunders} and \citet{Kolar+etal}; \citet{Bleecker} discusses jet bundles of connections on a principal bundle.}  For our purposes a Lagrangian is a scalar field on the space of 1-jets over some bundle of field configurations.  Note that this definition already builds in an important constraint: namely, that we are only considering first-order Lagrangian densities, i.e., Lagrangian densities that depend only on the first derivative of a field.

The space of possible derivative operators requires somewhat more work.  Fix a vector bundle $V\hookrightarrow E\xrightarrow{\pi} M$.  As we have already noted, given one derivative operator on a vector bundle, we may characterize (all) others using tensor fields.  In the Klein-Gordon case already described, this field may be written as a 1-form on $M$, $A_a$; in the case of a general vector bundle, it will be an endomorphism-valued 1-form.  But the derivative operators themselves are not tensor fields on $M$ nor on $E$.  They \emph{can} be understood as tensor fields over another bundle, canonically associated with $E$: namely, the frame bundle $\wp:FE\rightarrow M$ over $M$ associated with $E$.  The total space of this bundle consists in all bases for the fibers at each point of $E$; it inherits its bundle structure from $E$.  A section of $FE$ is a frame field, i.e., an assignment to each point of (an open subset of) $M$ of a basis for the fiber of $E$ over that point; the typical fibers are $G-$torsors for the general linear group on $V$, and thus $FE$ forms a principal bundle with structure group $GL(V)$.\footnote{A $G$-torsor is a smooth manifold, diffeomorphic to a Lie group $G$, with a free and transitive $G$ action.  Think of it as $G$ with its group structure forgotten.}  In the special case we considered in the previous section, of a bundle $\mathbb{C}\hookrightarrow E\xrightarrow{\pi} M$ of charged Klein-Gordon fields, the frame bundle may be thought of as the bundle of non-zero complex numbers.  This is because the non-zero complex numbers are precisely the possible bases for $\mathbb{C}$ conceived as a vector space, and they also correspond to the invertible linear transformations on $\mathbb{C}$.

The frame bundle is useful for present purposes because there is a one-to-one relationship between the derivative operators on $E$ and the \emph{principal connections} on $FE$, which we define as follows.  Let $\mathfrak{gl}(V)$ denote the Lie algebra associated with the general linear group on $V$, and let $ad:GL(V)\rightarrow GL(\mathfrak{gl}(V))$ be the \emph{adjoint representation} of $GL(V)$ on its own Lie algebra.  Then a principal connection is a Lie algebra valued one-form on $FE$, written $\omega^{\mathfrak{A}}{}_{\alpha}$,\footnote{Again, see \citet{WeatherallFBYMGR} for more on this notation.  Here the raised index $\mathfrak{A}$ indicates membership in a Lie algebra, and the lowered Greek index is a tangent index on $FE$.  Thus, $\omega^{\mathfrak{A}}{}_{\alpha}$ is a linear map from tangent vectors at a point $x\in FE$ to the Lie algebra $\mathfrak{gl}(V)$.} that is \emph{equivariant} in the sense that given any $g\in GL(V)$, any $x\in FE$, and any vector $\xi^{\alpha}$ at $x$, \begin{align}
\label{eq:equivariance}(\omega^{\mathfrak{A}}{}_{\alpha})_{|xg}(R_g)_*(\xi^{\alpha})=ad(g^{-1})^{\mathfrak{A}}{}_{\mathfrak{B}}(\omega^{\mathfrak{B}}{}_{\alpha})_{|x}\xi^{\alpha}.\end{align}
Here $(R_g)_*$ is the pushforward along the right action of $g$ on $FE$.  To understand this condition, note that a point $x$ in $FE$ may be thought of as a basis, and the right action of $g$ on $x$ may be thought of as implementing a change of basis, from $x$ to $xg$.  Thus, the condition encodes how the connection behaves under changes of basis.  Each principal connection on $FE$ uniquely determines a derivative operator on $E$, and all derivative operators on $E$ arise in this way.

We noted above that when the fibers of $E$ have further structure, such as the Hermitian inner product on the Klein-Gordon bundle, we often restrict attention to derivative operators that are compatible with that structure in the sense of making it constant.  We can transfer that condition to principal connections as follows.  (We will focus on the present example, though the construction proceeds similarly in other cases.)  First, observe that a Hermitian inner product determines a special class of frames, namely the \emph{orthonormal frames}.  The orthonormal frames of the Klein-Gordon bundle are just unit vectors, i.e., complex numbers of modulus 1. These form a submanifold of $FE$, and likewise a Lie subgroup of $GL(\mathbb{C})$, isomorphic to $U(1)$.  Thus we find, associated with $\pi:E\rightarrow M$, a principal sub-bundle of $FE$, $\wp_{|P}:P\rightarrow M$, with structure group $U(1)$.  We may define principal connections on this principal bundle as well; it is a well-known result that every principal connection on $P$ may be uniquely extended to a principal connection on $FE$ \citep{Kobayashi+Nomizu}.  It is precisely these connections---that is, the principal connections on $P$---that correspond to derivative operators on $E$ compatible with the inner product.  In what follows we will restrict attention to $P$.

We now turn to (first-order) Lagrangians for principal connections.  First, observe that every principal connection $\omega^{\mathfrak{A}}{}_{\alpha}$ is itself a section of a bundle whose fiber at each point $x$ consist of maps from $T_xP$ to $\mathfrak{g}$. We will call this bundle $\pi_C:\mathcal{C}\rightarrow P$.  This is the bundle of $\mathfrak{g}$-valued 1-forms on $P$; We can define it as the fiberwise tensor product bundle of the trivial bundle $\mathfrak{g}\times P\rightarrow P$ and the cotangent bundle $T_*P\rightarrow P$. (Note that $P$, the total space of a $U(1)$ bundle, is now the \emph{base space} for the new bundle $\mathcal{C}$.)  Not every section of this bundle is a principal connection, since not every section will satisfy the equivariance property in Eq.~\eqref{eq:equivariance}; in fact, the principal connections are precisely those sections of this bundle that do satisfy Eq.~\eqref{eq:equivariance}.

The first jet bundle of $\mathcal{C}$, $J^1\mathcal{C}\rightarrow \mathcal{C}\rightarrow P$, can be thought of as the bundle whose fibers, at each point of $\mathcal{C}$, consist of the affine space of ``first-derivatives'' of sections of $\mathcal{C}$ that pass through that point.  More precisely, we take $J^1\mathcal{C}$ to be the manifold of triples $(x,\omega,H_{(x,\omega)})$, where $H_{x,\omega}\subseteq T_{(x,\omega)}\mathcal{C}$ is such that the pushforward of $H_{x,\omega}$ along $\pi_{C}:C\rightarrow P$ is a bijective linear map.\footnote{Another way of thinking about this would be to define ``1-jets'' of connections at a point as equivalence classes of Lie algebra-valued 1-forms that agree on their first derivative at that point.  But ``agreement on first derivative'' in this context means having the same image under their pushforward, i.e., determining the same horizontal subspace at a point.  The approaches are equivalent, but we prefer to describe the geometric structure picked out by the equivalence class rather than work with the equivalences classes themselves.}  (We call such spaces ``horizontal subspaces'' at $(x,\omega)\in\mathcal{C}$, because they consist only of horizontal vectors, and they form a complement to the vertical space at $(x,\omega)$, i.e., the kernel of the pushforward along $\pi_C$.)  These spaces can be thought of as first derivatives of sections of $\mathcal{C}$ in the sense that they are the possible images of the tangent space of $P$ at $x$ under the pushforward along a section of $\mathcal{C}$, or, equivalently, they are the tangent spaces of sections of $\mathcal{C}$, encoding how a section changes, to first order, in each tangent direction at $x$.  Note that every principal connection $\omega^{\mathfrak{A}}{}_{\alpha}:P\rightarrow\mathcal{C}$ determines a section of $J^1\mathcal{C}\rightarrow P$, written $j^1\omega$, whose value at each point $x$ is given by $j^1\omega(x) = (\omega_{|x},\omega_*[T_xP])$, where $\omega_*:TP\rightarrow TC$ is the pushforward along $\omega^{\mathfrak{A}}{}_{\alpha}$.

It will be useful, for what follows, to see how the equivariance property of principal connections in Eq.~\eqref{eq:equivariance} imposes a condition on 1-jets of principal connections.  To do this, first note that the right action of $G$ on $P$ extends to actions on $\mathcal{C}$ and on $J^1\mathcal{C}$.  The right action on $\mathcal{C}$ can be written as $R^0_g:(x,(\omega^{\mathfrak{A}}{}_{\alpha})_{|x}))\mapsto (x\cdot g, ad(g^{-1})^{\mathfrak{A}}{}_{\mathfrak{B}}(\omega^{\mathfrak{B}}{}_{\alpha})_{|x} (R_{g^{-1}})_*)$, for any $g\in G$.  For convenience, we will write $\omega\cdot g$ as a shorthand for $ad(g^{-1})\omega_x (R_{g^{-1}})_*$.  Since the right action of $G$ on $P$ is smooth, this mapping is also automatically smooth.  It thus admits a differential (pushforward), which we can write $(R^0_g)_*:T\mathcal{C}\rightarrow T\mathcal{C}$.  This map takes tangent vectors at points $(x,\omega)$ to tangent vectors at points $(x\cdot g,\omega\cdot g)$.  We use this differential to define the right action on $J^1\mathcal{C}$, as $R^1_g:(x,\omega,H_{x,\omega})\mapsto (x\cdot g, \omega\cdot g,(R^0_g)_*[H])$, for every $g$.  Again, for convenience, we will write $H\cdot g$ as a shorthand for $(R^0_g)_*[H]$.  To see the value of these definitions,  observe that the equivariance property of principal connections implies that for any connection $\omega^{\mathfrak{A}}{}_{\alpha}$, any point $x$, and any group element $g\in G$, $j^1\omega(x\cdot g) = (\omega_{|x}\cdot g, \omega_*[T_xP]\cdot g) = j^1\omega(x)\cdot g$.  This is a necessary condition for a section of $J^1\mathcal{C}$ to be a principal connection.

Finally, a \emph{first-order Lagrangian for principal connections} is a smooth map $\mathcal{L}:J^1\mathcal{C}\rightarrow \mathbb{R}$, i.e., a scalar field on the first jet bundle over $\mathcal{C}$, that determines a well-defined function from $\hat{\Gamma}(P,C)$ to $C^{\infty}(M)$.  Here $\hat{\Gamma}(P,C)$ is the space of smooth, equivariant sections of $C$ -- that is, the space of principal connections -- and $C^{\infty}(M)$ is the space of smooth scalar fields on $M$.  This occurs precisely when $\mathcal{L}(x,\omega,H)=\mathcal{L}(x\cdot g,\omega\cdot g, H\cdot g)$.  Why?  The reason is that any smooth principal connection can be seen to determine a submanifold of $J^1\mathcal{C}$, given by $j^1\omega[P]$.  For $\mathcal{L}$ to determine a well defined function $\hat{\Gamma}(P,C)\rightarrow C^{\infty}(M)$, for every point $p\in M$, it must assign the same value to every point in $(\wp\circ\pi_{J^1\mathcal{C}})^{-1}[p]\cap j^1\omega[P]$, where $\pi_{J^1\mathcal{C}}$ is the projection from $J^1\mathcal{C}$ to $P$.  But if some point $(x,\omega_x,H)$ lies in $(\wp\circ\pi_{J^1\mathcal{C}})^{-1}[p]\cap j^1\omega[P]$, then all other points in $(\wp\circ\pi_{J^1\mathcal{C}})^{-1}[p]\cap j^1\omega[P]$ are of the form $(x\cdot g,\omega_x\cdot g,H\cdot g)$, for some (unique) $g\in G$ or other.  (This follow from the fact that $G$ acts freely and transitively on $P$.)  Since every covariant derivative operator on a vector bundle determines a unique principal connection on its frame bundle (and vice versa), it follows that first-order Lagrangians for principal connections can also be seen to determine well-defined maps from covariant derivative operators on vector bundles to smooth scalar fields on $M$.

Finally, for future reference, we note a special class of first-order Lagrangians for principal connections.  Every principal connection $\omega^{\mathfrak{A}}{}_{\alpha}$ on $P$ can be associated with a \emph{curvature tensor}, which is a tensor field on $P$ of the form $\Omega^{\mathfrak{A}}{}_{\alpha\beta}$, antisymmetric in $\alpha,\beta$, which is both equivariant, in a sense analogous to how connections are equivariant, and also horizontal, in the sense that for any vector $\xi^{\alpha}$ in the kernel of the pushforward along $\wp$, $\Omega^{\mathfrak{A}}{}_{\alpha\beta}\xi^{\beta}=\mathbf{0}$.\footnote{We said that the curvature is an endomorphism-valued two-form, whereas here it is presented as a Lie-algebra-valued two-form. The difference is important to track, but has a simple explanation: in a matrix representation of a Lie group on a vector space, Lie algebra elements are represented as general endomorphisms.  So the endomorphism-valued form noted above is just the curvature expressed as a field on $E$, whereas the curvature we consider here is expressed as a field on $FE$.}  The curvature tensor depends on a connection only to first order, in the sense that there exists, for every principal bundle, a well defined map $\hat{\Omega}$ from $J^1\mathcal{C}$ to the space of horizontal and equivariant Lie-algebra-valued two forms on $P$ with the property that, for any connection $\omega^{\mathfrak{A}}{}_{\alpha}$, $\hat{\Omega}(j^1\omega) = \Omega^{\mathfrak{A}}{}_{\alpha\beta}$, where $\Omega^{\mathfrak{A}}{}_{\alpha\beta}$ is the curvature associated with the connection $\omega^{\mathfrak{A}}{}_{\alpha}$.  A \emph{curvature Lagrangian} for principal connections is a Lagrangian for principal connections that can be written as $\mathcal{L} = K\circ \hat{\Omega}$, where $K$ is some smooth map from horizontal and equivariant Lie-algebra-valued two forms on $P$ to $\mathbb{R}$---that is, some way of building ``curvature scalars'' out of curvature tensors.  As an example, the Yang-Mills Lagrangian, which in general can be written as $k\Omega^{\mathfrak{A}}{}_{\alpha\beta}\Omega_{\mathfrak{A}}{}^{\alpha\beta}$, for some constant $k$, is a curvature Lagrangian; in the special case of electromagnetism, this becomes $F^{ab}F_{ab}$, where $F_{ab}$ is interpreted as the Faraday tensor.\footnote{The indices in the Yang-Mills action are raised/lowered with a combination of the Killing form, which is an invariant metric on the Lie algebra of any Lie group, and the spacetime metric, pulled back along the projection map from the principal bundle.  (This last construction makes sense because the lowered indices on $\Omega^{\mathfrak{A}}{}_{\alpha\beta}$ are horizontal.)  $\Omega^{\mathfrak{A}}{}_{\alpha\beta}$ becomes $F_{ab}$ for electromagnetism when we note that for the group $U(1)$, the associated Lie algebra is $\mathbb{R}$, and we use the fact that the curvature is horizontal to determine an action on vectors tangent to the base manifold, thereby exchanging the Greek indices for Latin ones.}

\subsection{Utiyama's Theorem}

We now return to the question of how dynamics are constrained by the structural posit that the Lagrangian should depend only on a vector bundle and connection, and no further structure (such as a preferred trivialization). As we have considered them so far, first-order Lagrangians for connections are required to be well-defined functions on connections, up to first-order.  But in principle, they could depend on other fields defined on a principal bundle, or they could depend on other structure, such as a preferred connection on the principal bundle.  We are interested in the special case of first-order Lagrangians that depend \emph{only} on the principal connection and principal bundle structure, and nothing else.  

We will make this idea precise with a symmetry condition.  Let $\wp:P\rightarrow M$ be a principal bundle.  A \emph{vertical principal bundle automorphism} is a diffeomorphism $\Psi:P\rightarrow P$ such that (1) $\wp = \wp\circ \Psi$ and (2) for all $x\in P$ and $g\in G$, $\Psi(x\cdot g) = \Psi(x)\cdot g$.  The first condition captures the idea that the diffeomorphism preserves the bundle structure (and acts trivially on the base manifold---hence the ``vertical''), while the second captures the idea that the diffeomorphism preserves the principal bundle structure, via the right $G$ action.\footnote{\label{fn:vertical} Why consider only vertical automorphisms?  In principle, we could consider all principal bundle automorphisms, and impose a still-stronger condition on Lagrangians below.  Utiyama's theorem would still follow.  But the fact is that the stronger condition is not needed.  Though see footnote \ref{fn:natDiffeos}.}  Since $\Psi$ is a diffeomorphism, we can define a pushforward map, which can be extended, in a straightforward way, to act on principal connections.  It is a standard result, then, that if $\omega^{\mathfrak{A}}{}_{\alpha}$ is a principal connection on $P$, then $\Psi_*(\omega^{\mathfrak{A}}{}_{\alpha})$ is also a principal connection on $P$.  In general, for non-Abelian $G$, $\omega^{\mathfrak{A}}{}_{\alpha}\neq \Psi_*(\omega^{\mathfrak{A}}{}_{\alpha})$; instead, we say that $(P,\omega^{\mathfrak{A}}{}_{\alpha})$ and $(P,\Psi_*(\omega^{\mathfrak{A}}{}_{\alpha}))$ are isomorphic.\footnote{This is entirely analogous to the general relativistic case, where for a metric $g_{ab}$ on a manifold $M$, and a diffeomorphism $\psi:M\rightarrow M$, in general $\psi_*(g_{ab})\neq g_{ab}$.  Nonetheless, the spacetimes $(M,g_{ab})$ and $(M,\psi_*(g_{ab}))$ are isomorphic \citep[c.f.][]{WeatherallHA}.}  That they are isomorphic means that they are associated to (or, arise as the frame bundles of) isomorphic vector bundles, carrying isomorphic derivative operators.

We will capture the idea that a first-order Lagrangian $\mathcal{L}$ for principal connections on a principal bundle $P\rightarrow M$ depends only on the connection (and principal bundle) via the condition that its action is invariant under principal bundle isomorphisms: that is, for any connection $\omega^{\mathfrak{A}}{}_{\alpha}$ and any vertical principal bundle automorphism $\Psi:P\rightarrow P$, $\mathcal{L}(j^1\omega)=\mathcal{L}(j^1(\Psi_*(\omega)))$.\footnote{This condition is very closely analogous to the standard condition in general relativity that a Langragian be ``diffeomorphism invariant'', which similarly captures a sense in which the Lagrangian has no implicit dependence on background structure.  Note that the condition does not quite say that $\mathcal{L}\circ\Psi_*=\mathcal{L}$, because not all points of $J^1\mathcal{C}$ coincide with principal connections.  Rather, the condition is equivalent to saying that $\hat{\mathcal{L}}\circ\Psi_*=\hat{\mathcal{L}}$, where $\hat{\mathcal{L}}$ is the map from connections to smooth scalar fields on $M$ induced by $\mathcal{L}$.}  Since $\omega^{\mathfrak{A}}{}_{\alpha}$ and $\Psi_*(\omega^{\mathfrak{A}}{}_{\alpha})$ are isomorphic, i.e., have the same structure as principal connections on a principal bundle, a Lagrangian that depends only on that principal connection structure should not be able to distinguish them, and thus should yield the same result when acting on both.  Conversely, this means that any Lagrangian satisfying the invariance condition cannot depend on any structure that is \emph{not} preserved by every principal connection isomorphism of this form.  In particular, it cannot depend on background structure such as a preferred section of $P$, or a preferred trivialization of any associated vector bundle, because that would not be preserved by maps of this sort.\footnote{Note that in many treatments, vertical principal bundle automorphisms are called ``gauge transformations'', and the invariance condition we have just described is called ``gauge invariance'' for Lagrangians.  We avoid that usage, however, because we think it obscures the sense in which this sort of invariance probes structural dependencies.  That is because in other contexts, gauge transformations are treated as ad hoc transformations with no clear geometric significance.  Exploring that relationship would take us too far afield, though see \citet[Ch. 3]{Bleecker} and \citet{WeatherallFBYMGR}.}

We now state Utiyama's theorem \citep[c.f.][Thm. 10.2.15]{Bleecker}.
\begin{thm}[Utiyama's Theorem] A first-order Lagrangian for principal connections on a principal bundle $P\rightarrow M$ is invariant under principal bundle isomorphisms only if it is a curvature Lagrangian.\footnote{The converse is also true, as long as one restricts attention to curvature Lagrangians that are ``adjoint invariant'', i.e., $K(ad(g)^{\mathfrak{A}}{}_{\mathfrak{B}}\Omega^{\mathfrak{B}}{}_{\alpha\beta})=K(\Omega^{\mathfrak{A}}{}_{\alpha\beta})$ for any $g\in G$, where $K$ is the map from Lie-algebra valued horizontal and equivariant two-forms introduced above.  But that direction does not matter for our argument here.}\end{thm}
\noindent In other words, any Lagrangian that depends only on the principal bundle and principal connection structure, to first order, must be a scalar function of the curvature associated with the connection.  Or, returning to the discussion in section \ref{GArevisited}, once we stipulate that the derivative operator acting on our fields is determined dynamically, and we require that that dynamics be given by extremizing a first-order Lagrangian that depends only on the bundle structure and derivative operator, with no further background structure, it follows that that Lagrangian must be a curvature Lagrangian.  

Utiyama's theorem by itself does not uniquely pick out the Yang-Mills Lagrangian, though as we have seen the Yang-Mills Lagrangian does count as a curvature Lagrangian. But when we add the additional constraint that the Lagrangian can depend only on structure we have already introduced, which includes the Killing metric and the spacetime metric but not other fields, we find that the Yang-Mills Lagrangian is, in a certain straightforward sense, the simplest possible non-trivial dynamics allowed.\footnote{By ``non-trivial'' we mean to rule out the trivial curvature Lagrangian, $\mathcal{L}=\mathbf{0}$.} This is because, as we have seen, the curvature is a Lie-algebra-valued two-form $\Omega^{\mathfrak{A}}{}_{\alpha\beta}$.  To construct a scalar from it, we must somehow contract the indices on this field.  This requires some other structure.  In the present case, though, the structure we have already introduced suffices, since we can take inner products with the metrics acting on the Lie algebra and horizontal indices, respectively.  Doing so produces the Yang-Mills Lagrangian.  In this sense, the Yang-Mills Lagrangian is the only non-trivial curvature Lagrangian (up to an overall constant prefactor) that is \emph{quadratic} in the curvature, which is the lowest order for which non-trivial curvature Lagrangians can be constructed, and which depends only on the other background structure already posited.\footnote{Recall fn. \ref{fn:vertical}. One way to make this uniqueness claim precise would be to consider a broader class of transformations to require ``dependence'' on only the metrics.  But doing this would require additional technical machinery that is not necessary for the point we wish to make. \label{fn:natDiffeos}.}

\section{Resolving the problems?}\label{criticsRevisited}

In order to see whether this new version of the gauge argument is justified, we will ask how it fares against Martin's three criticisms.

The first criticism was that the usual ways of justifying the move from global to local gauge invariance fall short.  We have not invoked either of the two problematic justifications (the one from local field theory, or the one from relativity) that Martin criticized.  We have instead used transformations that preserve various structures as a diagnostic for the dependencies of a Lagrangian.  Martin suggests that in the geometrical setting, there is not much of a distinction between global and local transformations \citep[p. 225-6]{Martin}.  He takes this to be grist for his mill. Any change of global trivialization is equally-well interpreted as mere changes of coordinatization of our fiber bundle, whether it is local or global.  Indeed, there is a strong sense in which whether such a transformation should even be classified as global or local is itself gauge-dependent---it depends on an initial choice of how to represent our fields.  If gauge-dependent facts are not physically significant, then the distinction between local and global should not be either \citep[c.f.][3.2.4]{Bleecker}.  So little weight should be put on the move from global to local transformations.  

We agree with these general statements, and especially the claim that the local/global distinction is gauge-dependent.  But note that in our version of the argument, there was still an important role played by the local/global distinction.  The reason was that we began by interpreting complex scalar fields as requiring a fixed global trivialization.  With that structure fixed, a global gauge transformation is one that is relatively constant, i.e., one that preserves the associated derivative operator.  We found that \emph{those} transformations preserve the Lagrangian, whereas more general changes of trivialization, which include the merely local transformations and do not necessarily preserve the derivative operator, do not preserve the Lagrangian.  This turned out to be a particularly clear way to show that the Lagrangian depended on a derivative operator but not a global trivialization, because the Lagrangian was invariant under some, but not all, changes of trivialization---namely, the ones that preserved the derivative operator.  In other words, to return to the analogy with the Weyl-Stein argument, the difference between local and global gauge transformations is analogous to the difference between Leibnizian and Galilean transformations, which both play a role in identifying a derivative operator as the intermediate structure between a global trivialization and a bare bundle with metrics needed for physics.

The second criticism was that even after imposing local gauge invariance, the Lagrangian is not uniquely determined.  One needs further constraints, for example, concerning Lorentz invariance and simplicity to figure out which gauge invariant terms to add and which to exclude.  On the geometrical account, we do not claim that the gauge argument uniquely determines the interaction Lagrangian.  We merely conclude with certain constraints on the form the Lagrangian can take, based on what structure it can depend on.  The Yang-Mills Lagrangian stands out as the only quadratic curvature Lagrangian depending only on the allowed structure. (Note that this structural restriction plays the role of Lorentz invariance, since we take inner products only with the spacetime metrics.)   More work can be done by empirical testing, for example, to determine the equations of motions for the fields, which then constrains the Lagrangian even further.  It is a combination of conceptual and empirical considerations that lead us to the final Lagrangian and a full blown field theory.  We need not be worried about this; all we have noticed is that constructing a physical theory involves multiple considerations, and that the geometrical formulation of electromagnetism gives us only some constraints on the final form of the theory.

There is another perspective on this issue, though, which deserves further attention (in future work).  The Lagrangian we arrive at, by beginning with the Klein-Gordon Lagrangian, recognizing that it depends on a derivative operator (but not a global trivialization), and then introducing dynamics for that derivative operator via Utiyama's theorem, is often described as a \emph{minimally coupled} Lagrangian.  In principle other terms could appear; our version of the gauge argument excludes them only because there is no reason to introduce them:~they neither arise from the Klein-Gordon equation relative to a flat derivative operator, nor do they arise from Utiyama's theorem. They would have to be added by some other argument (even though they would not depend on additional structure).  Recent work on minimal coupling in the context of a Lorentzian metric, however, suggests that there may be other reasons to prefer minimally coupled equations, such as that they have nice causal properties \citep{Weatherall+MarchNT,March+WeatherallJMP}.  Extending that work to include equations that depend only on a principal connection could be fruitful.

The third criticism was that the reality of the field $F_{ab}$ was essentially ``put in by hand", even though it was supposed to be one of the great things that the gauge argument revealed to us.  On the geometrical account, we started with some geometrical facts about our principal bundle including a connection and its curvature.  It should be no surprise that these lead to physically interesting interactions and that representing these geometrical objects by bringing them down onto spacetime gives us physically significant fields.  So yes, the fields were in some sense put in by hand in the geometrical structure that we began with.  But that is not problematic. The existence of the fields is not something that we are supposed to conclude from the argument.  In fact, the fields, or more specifically the connection and curvature that give rise to them, are what fuel the gauge argument in the first place.

Recall that the initial philosophical puzzlement over the gauge argument was that it appeared to `derive something from nothing'.  We started with the requirement of gauge invariance, which just signifies a descriptive redundancy in the mathematical formalism, and we concluded with a new Lagrangian and the fact that the photon is massless (and some would add the existence of new fields).  On the geometrical account, we see that the `nothing' we started with was much more than nothing, and we can construe the `something' that we conclude with much more modestly.  On this view, gauge invariance is seen as more than a descriptive or formal redundancy.  It arises from a formulation in terms of fiber bundles, where there are physically significant facts concerning the connection and curvature of the space in which fields are valued.  Of course there are different ways of representing those facts in different coordinate systems, or in other words through different sections.  But certainly the geometry that we start with is much more than nothing.  What we concluded from the gauge argument was merely that there are certain constraints on the final form of the Lagrangian for the theory.  We did not claim that gauge invariance was the only factor that uniquely leads to a full blown interacting field theory.  Even so, many of the initial worries about the philosophical status of the gauge argument seem to dissolve.  On this view, we do not derive something from nothing, but we do see a sense in which the mathematical formalism of Yang-Mills theory and the Lagrangian describing that theory mutually constrain one another.\footnote{We can see, here, an extension of Weatherall's ``puzzleball conjecture'' to Yang-Mills theory \citep{WeatherallPuzzleball}.}

\section*{Acknowledgments}
This paper is based upon work supported by the National Science Foundation under Grant Nos. 1331126 and 2419967 (Weatherall) and Graduate Research Fellowship No. DGE-1321846 (Feintzeig).  Preliminary versions of this paper were presented at the second IPP Conference on the Conceptual and Mathematical Foundations of Physics in March 2014, hosted at UC Irvine; as part of a masterclass in Philosophy of Physics at Cambridge University in 2016; and as a pair of talks to the Foundations seminar at the Black Hole Initiative in 2021. we are grateful to the audience for comments and questions.  We are also grateful to Jeff Barrett, India Bhalla-Ladd, Jeremy Butterfield, Sam Fletcher, Henrique Gomes, David Malament, Eleanor March, and Sarita Rosenstock for helpful conversations related to this material.

\bibliographystyle{elsarticle-harv}
\bibliography{gaugeArg}

\end{document}